\documentclass[aps,preprint]{revtex4-1}
\usepackage{graphicx}
\usepackage{amsmath}
\usepackage{makeidx}
\usepackage{amsfonts}
\usepackage{amssymb}
\usepackage{color,soul}

\begin{document}

\title{Circuit topology of linear polymers: a statistical mechanical treatment}

\author{Alireza Mashaghi$^a$, Abolfazl Ramezanpour$^b$}
\affiliation{$^a$Harvard Medical School, Harvard University, Boston, Massachusetts, USA}
\affiliation{$^b$Department of Physics, University of Neyshabur, Neyshabur, Iran}

\begin{abstract}

Circuit topology refers to the arrangement of interactions between objects belonging to a linearly ordered object set. Linearly ordered set of objects are common in nature and occur in a wide range of applications in economics, computer science, social science and chemical synthesis. Examples include linear bio-polymers, linear signaling pathways in cells as well as topological sorts appearing in project management. Using a statistical mechanical treatment, we study circuit topology landscapes of linear polymer chains with intra-chain contacts as a prototype of linearly sorted objects with interactions.  We find generic features of the topological space and study the statistical properties of the space under the most basic constraints on the occupancy of arrangements and topological interactions. We observe that a set of correlated contact sites (a sector) could nontrivially influence the entropy of circuits as the number of involved sites increases. Finally, we discuss how constraints can be inferred from the information provided by local contact distributions in presence of a sector. 

\end{abstract}

\maketitle

\section{Introduction}\label{S0}

\begin{figure}
\includegraphics[width=10cm]{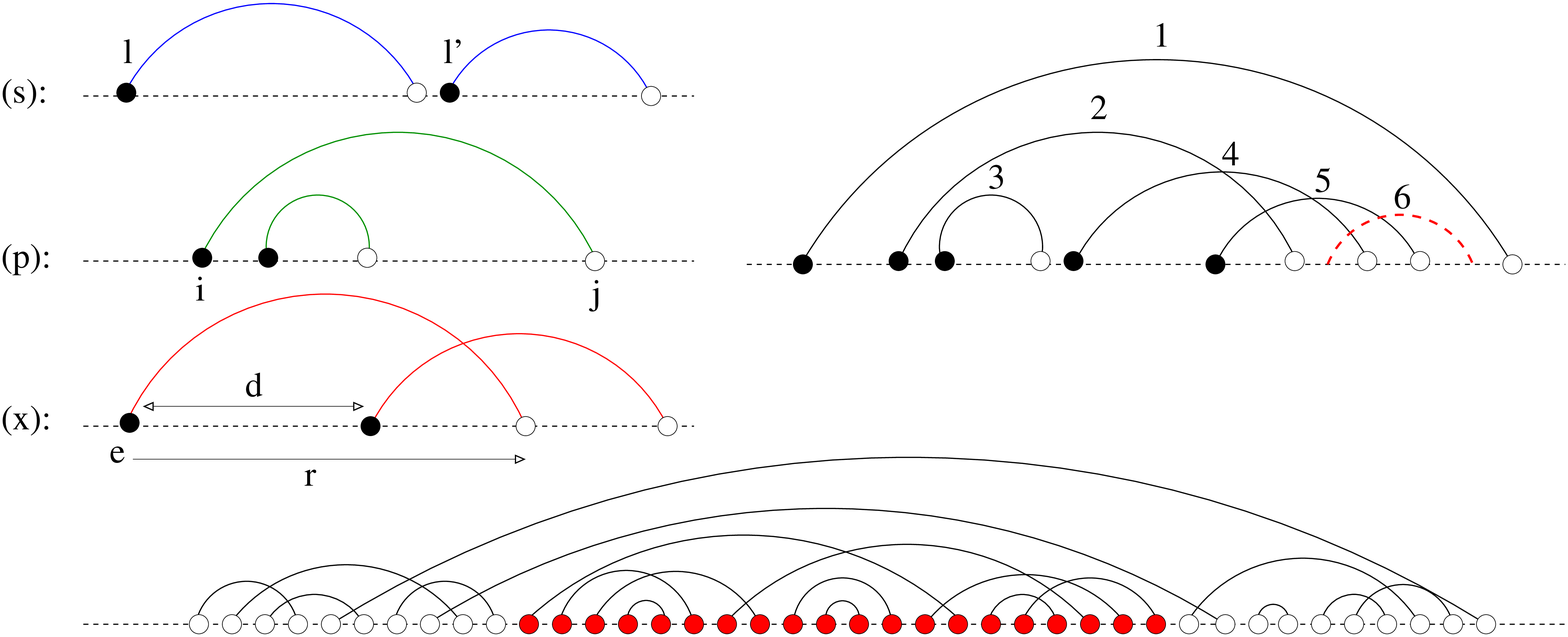} 
\caption{(top-left) Arrangements of two links $l,l'$ in series (s), parallel (p), and cross (x) states. A link can be represented by its endpoints $(i,j)$, or by its first endpoint and length $(e,r)$. Distance $d$ of two links is the separation of their first endpoints. (top-right) Links labeled according to their first endpoints to show the possible valid configurations for link $l=6$ given the links $l'=1,\dots,5$. (bottom) A sample link configuration in presence of a regular hard sector in the middle of the chain.}\label{f1}
\end{figure}

Molecular chains with intra-molecular contacts are topologically diverse. This is true even in the absence of branching and knot formation and is rooted in the variety of contact arrangements available to the chain (Fig. \ref{f1}, left). Circuit topology formalizes this notion using discrete mathematics and sets the stage for classification as well as functional and evolutionary analysis of (bio-)molecular chains based on their structural topology \cite{MWT-structure-2014}. Two chains with the same circuit topology may differ in total length and length of inter-contact segments but have equal number of contacts with identical contact arrangements. 

Circuit topology of a chain is a determinant of its function and dynamics and is in turn determined by the physico-chemical properties of the chain and its environment \cite{MWT-structure-2014, MTM-pccp-2014}. For example, folding rate of an isolated chain correlates with the number of contact pairs in parallel arrangement \cite{MTM-pccp-2014}. The topology influences whether distinct intermediate states are visited during folding and unfolding. Here crossed contacts are found to be the key determinants \cite{MTM-pccp-2014}. On the other hand, intrinsic and extrinsic factors determine the topological diversity of biomolecules. It is well established that the distribution of positively charged residues is a key determinant of membrane protein topology \cite{H-embo-1986}. When folding of a chain occurs while the chain is sequentially synthesized, certain topologies might be kinetically populated. Evolutionary constraints also select for certain topologies with desired stability and functionality \cite{H-nature-2006}. Slow folding chains are prone to unwanted interactions and aggregation and thus are disfavored. The constraints, such as those discussed above, are not represented in the free energy landscape of the chain, which maps conformations of isolated chains to their corresponding free energies.

Chain models with contacts have served as prototypes in theories of biomolecular chains and in particular RNA and protein folding and can be used to study circuit topology. Because the length of the chain is irrelevant in a topological treatment, the model can be further simplified by considering only the contacts and setting the length of every chain segment to unity. Contacts can be displayed as links between the contact sites. The chain will then be modeled by a connected graph in which the nodes correspond to the contact sites, the ordered sequence of contact sites corresponds to the polymer backbone and the remaining links represent the intra-chain contacts. The latter forms a perfect matching of the graph. This graph representation of the chain shrinks the conformational space of the chain to a topological space. 

Studying the topological diversity is technically challenging. The problem of sampling from the exponentially large space of contact configurations (perfect matchings) could be very time consuming for disordered and frustrated energy functions. An efficient way of sampling from such energy landscapes in sparsely (weakly) interacting systems is provided by the cavity method of statistical physics, relying on the Bethe approximation \cite{MGV-epl-1986,MP-epjb-2001,MM-book-2009}. The recursive and local nature of these equations are exploited in approximate message-passing algorithms that have proven useful in the study of random constraint satisfaction and optimization problems \cite{MZ-pre-2002,MPZ-science-2002,ZM-jstat-2006,MS-jstat-2006,RZ-pre-2012,BBCZ-jstat-2008}.

In this article, we discuss the statistical mechanical properties of a single chain that forms intra-chain binary contacts in the context of circuit topology. We use the Bethe approximation to characterize the space of contact (link) configurations assuming an energy function of two-link interactions depending on their relative position in the polymer chain. We illustrate the constraints imposed by the energy function on the configuration space, and obtain the one-link and two-link probability distributions to see how the links are organized by changing the relevant parameters in the energy function. We also obtain the entropy (logarithm of the number of contact configurations) in terms of the two-link densities in the energy function of the system. We will see how a subset of correlated contact sites identified by a sector \cite{HRLR-cell-2009} affects the statistical properties of the chain. In particular, for specific frustrating energy functions the entropy displays a maximum for sector sizes close to half the number of contact sites.    
Finally, given the one-link and two-link data from structured link configurations, we try to reconstruct the energy function that statistically describes the observed data. Using this information, we can recover the contact sites of regular sectors of different sizes with an accuracy that approaches one as the number of sector sites increases.

\section{Definitions}\label{S1}
A chain with $M$ contacts is represented with a graph containing $M$ links with endpoints $e_l=(i_l,j_l)$ labeled by $l=1,\dots,M$ with $i_l,j_l \in \{1,\dots,2M\}$. The chain is directed from left to right $C=\{1,2,\dots,2M-1,2M\}$. A link configuration is defined by $\mathsf{L}=\{(i_l,j_l)|l=1,\dots,M\}$ where $i_l \neq j_l$ and links are not directed $(i_l,j_l) \equiv (j_l,i_l)$. To any pair of links one of the three states may be assigned with respect to the backbone chain $C$:  parallel ($p$), series ($s$), or cross ($x$), see Fig. \ref{f1}.  We will study topologically different link configurations represented by an $M\times M$ matrix $\mathsf{A}_{l,l'} \in \{p,s,x\}$. Note that physical length is not relevant to our study and thus we do not care if the simplified model with the presented length profile has a physical 3D realization or not. 

Consider the perfect matchings of the $2M$ nodes $i=1,\dots,2M$ on chain $C$ where a perfect matching configuration is defined by $\mathsf{M}=\{c_{ij}=0,1|\sum_{j\ne i} c_{ij}=1, \forall i\}$. Each perfect matching represents a class of topologically equivalent link configurations related by a permutation of the link labels. Note that in a perfect matching labels are assigned only to the endpoints whereas in a link configuration both the endpoints and the links are labeled. The number of perfect matchings is $\frac{(2M)!}{2^MM!}=(2M-1)!!=(2M-1)\times(2M-3)\dots \times 1$ and for each one there are $M!$ ways of labeling the links. In other words there are $M!$ matrices $\mathsf{A}$ representing the set of topologically equivalent link configurations. 

A link configuration of $M$ links is composed of $N=M(M-1)/2$ link pairs that can be classified into three disjoint subsets of size $N_p,N_s,N_x$ depending on their states $p,s,x$. The links can have an arbitrary labeling; one special case is to order the links from left to right according to their first endpoints. At some point in this paper, we will consider structured link configurations with sectors; a sector is identified by an arbitrary set of endpoints that remain (possibly differently) connected in different link configurations; we say a sector is hard if any connection between the sector sites and the other sites is forbidden. Figure \ref{f1} illustrates the definitions and notations used throughout this paper.

\section{Characterizing the configuration matrix}\label{S2}
Given a perfect matching and a link labeling, one can easily construct the unique matrix $\mathsf{A}$; the endpoints $(i_l,j_l)$ and $(i_{l'},j_{l'})$ are enough to identify the element $\mathsf{A}_{l,l'}$. On the other hand, given a matrix $\mathsf{A}$ and a labeling, one can find the unique matching configuration corresponding to the matrix (if there exists) by solving the following constraint satisfaction problem
\begin{align}
\mathbb{I}(\mathsf{A})=\sum_{\mathbf{e}}  \prod_{l<l'} \delta_{e_l\neq e_{l'}} \prod_{l<l'}\delta_{\mathsf{A}_{l,l'},\mathsf{q}(e_l,e_{l'})}.
\end{align}
In words, we find the endpoints $e_l=(i_l,j_l)$ that make a perfect matching and are consistent with the matrix $\mathsf{A}$. 
Here $\delta_{e_l\neq e_{l'}}=1$ if $e_l$ and $e_{l'}$ represent two disjoint links with different endpoints, otherwise it is zero. And $\mathsf{q}(e_l,e_{l'})\in \{p,s,x\}$ returns the state of links $l$ and $l'$ given their endpoints. 
The indicator function $\mathbb{I}(\mathsf{A})$ represents the constraints on the valid matrices; $\mathbb{I}(\mathsf{A})=1$ if $\mathsf{A}$ is a valid matrix, otherwise it is zero. It seems that the constraints on the matrix elements $\mathsf{A}_{ll'}$ can not be expressed in a local way; one needs to consider the constraints imposed on any two matrix elements, three elements, and more. From a computational point of view, the problem of deciding on the validity of an arbitrary configuration matrix could be hard, but we will see that at least for a class of ordered matrices this problem is easy. 

Suppose the links are ordered according to their first endpoints, that is $i_{l'}<i_l$  if $l'<l$. We assume that $i_l <j_l$ for any link $l$.
Given matrix $\mathsf{A}$, we add the links $1,2,3,\dots$ one by one (see Fig. \ref{f1}, right) to find the matching configuration.   
In step $l$ we add link $l$ and determine the link configuration just by looking at the matrix elements $\mathsf{A}_{l,l'}$ for $l'<l$. We have to determine the relative position of the endpoints $(i_l,j_l)$ with respect to the $\{(i_{l'},j_{l'})|l'<l\}$. Clearly $i_l > i_{l'}$ for all $l'<l$ and we need to consider only the other endpoints $j_{l'}$. The relative position of these endpoints have already been determined in the previous steps of the process, say $j_{1}<j_{2}<\cdots j_{l-1}$. Then we group the previous links according to the matrix elements: $g_p=\{l'|\mathsf{A}_{l,l'}=p\}$, with $g_s$ and $g_x$ defined in a similar way.
Now it is clear that
\begin{align}
\begin{cases}
i_l<j_{l'}, \hskip0.1cm j_l<j_{l'} &\mbox{if } l' \in g_p,\\
i_l>j_{l'}, \hskip0.1cm j_l>j_{l'} &\mbox{if } l' \in g_s,\\ 
i_l<j_{l'}, \hskip0.1cm j_l>j_{l'} &\mbox{if } l' \in g_x. 
\end{cases}
\end{align}
This defines the relative position of link $l$ given links $l'<l$.

‍The previous paragraph somehow defines the constrains between the matrix elements $\mathsf{A}_{ll'}$. As before suppose the links are ordered according to their first endpoint, and we are to add link $l$ given the configuration of links $1,2,\dots,l-1$. This means that in matrix $\mathsf{A}$ we are at row $l$ and we want to specify the valid matrix configurations in that row $\{\mathsf{A}_{ll'}|l'<l\}$. The above arguments say that these elements are constrained to the following configurations:
\begin{align}
j_{1}  \dots j_{a} \hskip2mm \mathbf{i}_{l} \hskip2mm  j_{a+1}\dots  j_{b} \hskip2mm  \mathbf{j}_{l} \hskip2mm j_{b+1} \dots j_{l-1}
\end{align}
The endpoints $j_{l'}$ that happen before $i_l$ belong to the links of group $g_s$, those that happen after $j_l$ belong to the links of group $g_p$, and the middle ones belong to the links of group $g_x$.  In short, the matrix elements in rows $1,2,\dots,l-1$ define the set of possible matrix elements in row $l$.

\section{Characterizing the matching configurations}\label{S3}
\begin{figure}
\includegraphics[width=15cm]{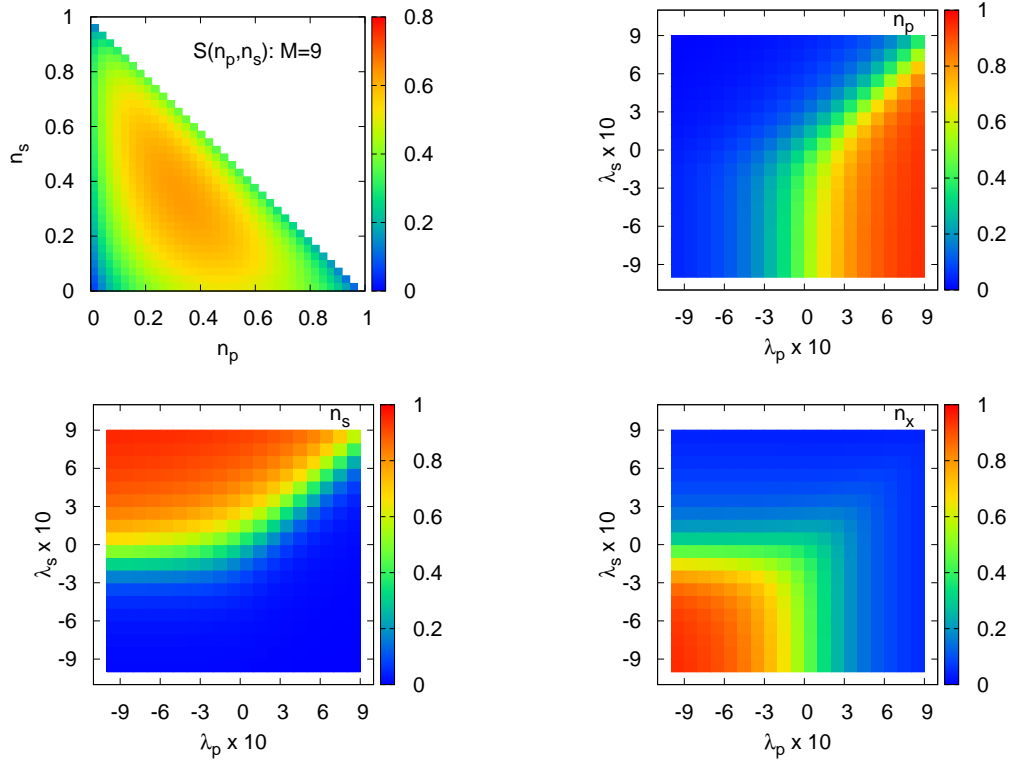} 
\caption{The entropy $S=\frac{1}{M\ln M}\ln \mathcal{N}$ vs the two-link densities $n_{p,s}=N_{p,s}/N$, and the two-link densities vs the energy parameters $\lambda_{p,s}$ for $\lambda_x=0$, obtained by an exhaustive enumeration algorithm for $M=9$ links. $\mathcal{N}$ is the number of configurations.}\label{f2}
\end{figure}

Consider matching configurations with a given number $N_p,N_s,N_x$ of link pairs $(l,l')$ of type $p,s,x$, respectively.  
We take $N=N_p+N_s+N_x=M(M-1)/2$ and define the energy function $E(n_p,n_s,n_x)=-M\ln M ( \lambda_p n_p+\lambda_s n_s+\lambda_x n_x) $ with densities $n_{p,s,x}=N_{p,s,x}/N$. The factor $M\ln M$ is chosen to have the same scaling for the energy and the leading term of the entropy function $S(n_p,n_s,n_x)$; we recall that the total number of link configurations scales as $e^{M\ln M}$ for large $M$.
Figure \ref{f2} shows the exact entropy we obtain for a small number of links; the entropy goes to zero for the all-$(p,s,x)$ configurations in the corners of the entropy plot, gets larger values when two types of contacts are allowed, and finally takes its maximum value for the neutral choice of the energy parameters $\lambda_{p,s,x}=0$, where $n_{p,s,x}=1/3$. In the figure we also see how the energy parameters control the two-link densities $n_{p,s,x}$; increasing $\lambda_{p,s,x}$ increases the probability of having a configuration with more of the corresponding type of contact.   
Note that the entropy distribution is broader in the $n_s$ direction and approaches a nonzero value for $n_s\to 0$. We observe the same behaviors for larger number of links using the following approximate algorithm.

Let us represent a perfect matching by the set of endpoints $\mathbf{e}=\{e_l=(i_l,j_l)|l=1,\dots,M\}$ assigned to the $M$ link variables, such that for any two different links $e_l\neq e_{l'}$. Then, we consider the following probability measure in the space of link configurations     
\begin{align}
\mu(\mathbf{e}) \propto \prod_{l<l'} \left( \delta_{e_l\neq e_{l'}} e^{ \tilde{\lambda}_p\delta_{\mathsf{q}(e_l;e_{l'}),p}+\tilde{\lambda}_s\delta_{\mathsf{q}(e_l;e_{l'}),s}+\tilde{\lambda}_x\delta_{\mathsf{q}(e_l;e_{l'}),x} } \right),
\end{align}
with $\tilde{\lambda}_{p,s,x}=2\frac{\ln M}{M-1}\lambda_{p,s,x}$.

We will compute the local marginals $\mu_l(e_l), \mu_{ll'}(e_l,e_{l'}),\dots$ of the $\mu(\mathbf{e})$  within the Bethe approximation. Moreover, we assume the relevant link configurations are organized in a simple and connected region of the configuration space; this is called the replica symmetric approximation \cite{MM-book-2009}. To this end, we need to write the recursive equations for the cavity marginals $\mu_{l\to l'}(e_l)$ of having endpoints $e_l$ for link $l$ in the absence of link $l'$,    
\begin{align}\label{bp}
\mu_{l\to l'}(e_l) \propto \prod_{l'' \ne l,l'} \left( \sum_{e_{l''} \ne e_l} e^{\tilde{\lambda}_p\delta_{\mathsf{q}(e_l;e_{l''}),p}+\tilde{\lambda}_s\delta_{\mathsf{q}(e_l;e_{l''}),s}+\tilde{\lambda}_x\delta_{\mathsf{q}(e_l;e_{l''}),x}} \mu_{l'' \to l}(e_{l''})\right).
\end{align}
These are the so-called belief propagation (BP) equations \cite{MM-book-2009,KFL-inform-2001}. Note that for large $M$ some of the cavity marginals $\mu_{l\to l'}(e_l)$ could take very small values increasing the numerical errors. To get around this problem, one can instead work with the cavity fields $h_{l\to l'}(e_l)\equiv \ln \left(\frac{\mu_{l\to l'}(e_l)}{\mu_{l\to l'}(e_0)}\right)$ with respect to a reference link variable $e_0$. 
   
The above equations can be solved by iteration starting from random initial messages. The fixed-point cavity marginals are enough to compute the interesting average quantities $\langle n_{p,s,x}\rangle$, and the entropy $S(n_p,n_s,n_x)$ as described in more details in Appendix \ref{BP-app}. In addition, we explain another efficient representation of the problem working with the matching variables $c_{ij}\in \{0,1\}$. In Appendix \ref{figs-app}, we give the parameter values for which the BP algorithm converges; the maximum entropy region around $\lambda_{p,s,x}=0$ is surrounded by a region that the algorithm does not converge, but still the entropy has a significant value. This could happen if strong correlations impose a more complicated organization of the relevant link configurations in the configuration space \cite{MM-book-2009}. 

\begin{figure}
\includegraphics[width=15cm]{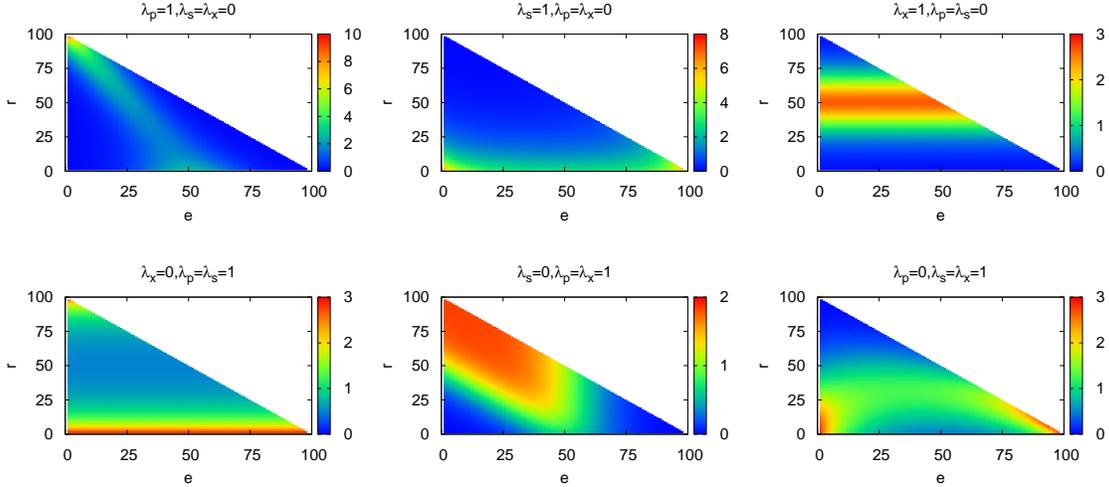} 
\caption{One-link distribution (more precisely $M(2M-1)\mu(e,r)$) obtained by the Bethe approximation for $M=50$ links and different energy parameters $\lambda_{p,s,x}$. Here $\mu(e,r)$ is the probability of having a link with the first endpoint $e$ and length $r$. The one-link distribution is uniform and thus trivial for $\lambda_{p,s,x}=0$ (not shown here). (top-left) Giving more weight to parallel two-links results to long links with first endpoints concentrated in the beginning and the first half of the chain. (top-middle) Giving more weight to series two-links results to short links with first endpoints nearly uniformly distributed along the chain. (top-right) Giving more weight to cross two-links results to links that are anywhere, but are of a particular length. (bottom-left) Giving less weight to cross two-links leads to a mixture of short and long links distributed uniformly along the chain.
(bottom-middle) Giving less weight to series two-links leads to links with a broad range of lengths, and starting mostly at the beginning of the chain.
(bottom-left) Giving less weight to parallel two-links leads to nearly short links concentrated mostly in the beginning and end of the chain.}\label{f3}
\end{figure}

\begin{figure}
\includegraphics[width=16cm,height=4cm]{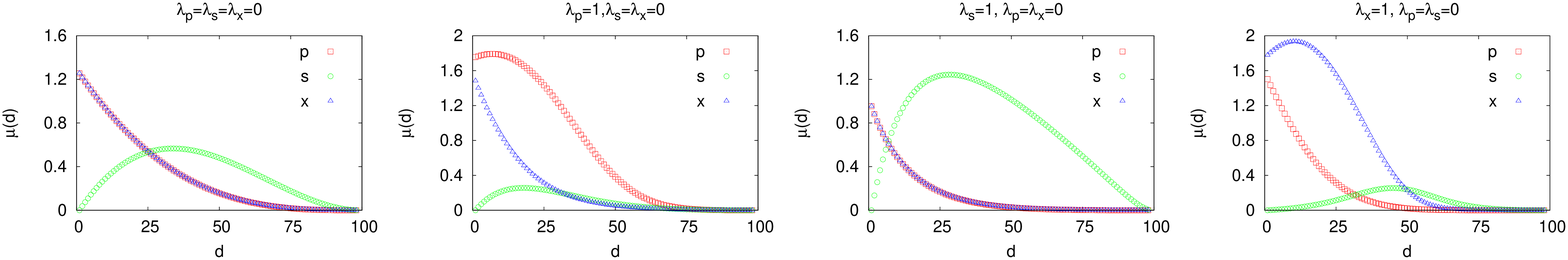} 
\caption{Two-link distance distribution $\mu_{ll'}(d)$ (multiplied by a constant to make it of order one) for different energy parameters $\lambda_{p,s,x}$ and different types of two-links $(p,s,x)$, obtained by the Bethe approximation for $M=50$ links. Here $\mu_{ll',q}(d)$ is the probability of finding two links of type $q=p,s,x$ at distance $d$ (separation of the first endpoints) from each other.}\label{f4}
\end{figure}

Figures \ref{f3}, and \ref{f4} display some typical one-link and two-link distributions obtained by the Bethe approximation. We see the structural properties of the link configurations change considerably around the origin of the parameter space $\lambda_{p,s,x}=0$, where the one-link probability distribution $\mu_l(e,r)$ is uniform.  In particular, the one-link distributions in Fig. \ref{f3} show that for $(\lambda_p=0,\lambda_{s,x}=1)$ we have short links that are mostly concentrated at the beginning and at the end of the chain in two communities. The other cases shown in the figure correspond to simpler structures dominated by one type $p,s,x$, or a superposition of two types. The two-link distance distributions $\mu_{ll'}(d)$ in Fig. \ref{f4} show that, as expected, there is always a nonzero length scale $d_s^*$ separating two links that are in series. However, starting from the neutral parameter values $\lambda_{p,s,x}=0$, the distance $d_s^*$ behaves differently 
by increasing the number of parallel or crossing two-links. In Appendix \ref{figs-app} we give the link distributions for more instances of the parameters along with a comparison of the approximate and exact data for small number of links. 

\section{Sectors: Implications and inference}\label{S4}
Let us consider a hard sector where any link connects either two sites inside or outside the sector set. The sector is defined by an arbitrary subset of contact sites $\mathcal{S}=\{i_1,\dots,i_{|\mathcal{S}|}\}$ that could be distributed randomly or regularly, and energy parameters $(\lambda_{p,s,x}^{\mathcal{S}},\lambda_{p,s,x}^{\mathcal{S}\bar{\mathcal{S}}},\lambda_{p,s,x}^{\bar{\mathcal{S}}})$. These parameters specify the relative importance of two-link arrangements for two links in the sector $(\lambda_{p,s,x}^{\mathcal{S}})$, one link in the sector and the other not in the sector $(\lambda_{p,s,x}^{\mathcal{S}\bar{\mathcal{S}}})$, and two links not in the sector $(\lambda_{p,s,x}^{\bar{\mathcal{S}}})$. Using this information, we can find an estimation of the entropy and other statistical properties of the chain by the BP equations given above; but, now an endpoint inside the sector can be connected only to another one in the same sector. In this section, we mainly focus on the inverse problem of inferring the sector from an appropriate set of observation data. Obviously, we first need to solve the froward problem of computing the expectation values of the relevant quantities, as explained in the previous section.  

\begin{figure}
\includegraphics[width=15cm]{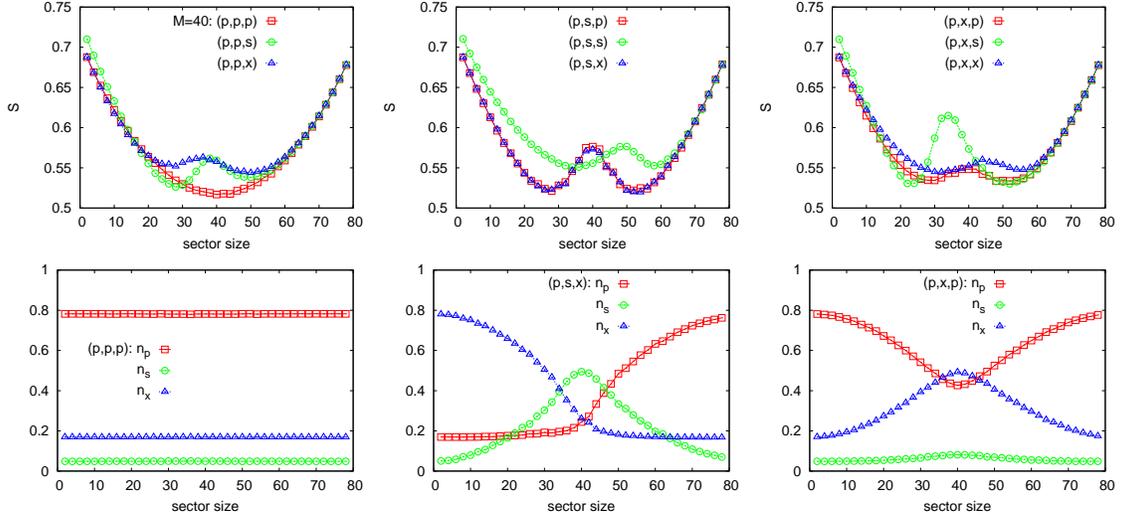} 
\caption{ The entropy $S$ and link densities $n_{p,s,x}$ vs the sector size obtained by the Bethe approximation for fixed energy parameters $(\lambda_{p,s,x}^{\mathcal{S}},\lambda_{p,s,x}^{\mathcal{S}\bar{\mathcal{S}}},\lambda_{p,s,x}^{\bar{\mathcal{S}}})$ with $M=40$ links. Here $\mathcal{S}$ stands for two links in sector, $\bar{\mathcal{S}}$ for two links not in sector, and $\mathcal{S}\bar{\mathcal{S}}$ for one link in sector and the other not in sector. To shorten the notation, we use for example $(p,x,s)$ to show $(\lambda_{p}^{\mathcal{S}}=1,\lambda_{s,x}^{\mathcal{S}}=0)$, $(\lambda_{x}^{\mathcal{S}\bar{\mathcal{S}}}=1,\lambda_{p,s}^{\mathcal{S}\bar{\mathcal{S}}}=0)$, and $(\lambda_{s}^{\bar{\mathcal{S}}}=1,\lambda_{p,x}^{\bar{\mathcal{S}}}=0)$ highlighting only the nonzero parameters. The data are averaged over $100$ realizations of randomly selected sector sites with relative errorbars of order $0.01$.}\label{f5}
\end{figure}

Figure \ref{f5} shows the number (entropy) of link configurations and two-link densities $n_{p,s,x}$ vs the number of randomly selected sector sites (sector size $|\mathcal{S}|$) for fixed energy parameters. Here, we are giving more weight to parallel two-links inside the sector and vary the other energy parameters. The entropy is, of course, larger for very small or large sector sizes than for intermediate sizes. However, depending on the energy parameters, the entropy could display a local maximum for sector sizes around $L/2$. In the same region, we observe convergence problems in the BP algorithm signaling the presence of strongly correlated link variables. Note that the number of forbidden link configurations increases by the size of sector. On the other hand, when the energy parameters $(\lambda_{p,s,x}^{\mathcal{S}},\lambda_{p,s,x}^{\mathcal{S}\bar{\mathcal{S}}},\lambda_{p,s,x}^{\bar{\mathcal{S}}})$ are different, new link configurations could appear as the sector size increases. The local maximum in the entropy is observed if the number of new configurations dominates the number of forbidden ones. When this happens, as the figure shows, the differences in the two-link densities become smaller making the system closer to the absolute maximum entropy point, where $n_{p,s,x}=1/3$.    

The energy function we considered in the above sections was indeed devised to explore the configuration space for different densities $n_{p,s,x}$. To have more control on the structure of the link configurations we have to consider more general energy functions.
Suppose we are given the average numbers $M^*(r)$ of links of length $r$, and numbers $N_{p,s,x}^*(d)$ of link pairs of type $p,s,x$ with distance $d$ between their first endpoints. From the maximum entropy principle \cite{BR-arxiv-2007}, the right energy function to model the system is
\begin{align}
E(\mathbf{e})=\sum_r \lambda(r) \left(\sum_{l}\delta_{j_l-i_l,r}\right)+\sum_{q=p,s,x}\sum_{d} \lambda_q(d)\left(\sum_{l<l'}\delta_{\mathsf{q}(e_l,e_{l'}),q}\delta_{|i_l-i_{l'}|,d}\right).
\end{align}
Given the parameters $\lambda(r)$ and $\lambda_{p,s,x}(d)$, we can use Bethe approximation as before to compute the averages $\langle M(r) \rangle$  and $\langle N_{p,s,x}(d) \rangle$. In the inverse problem, we are given the average numbers, and look for the energy parameters describing the data \cite{WWSHH-pnas-2009,PZS-plos-2011,PZW-pnas-2011}.

\begin{figure}
\includegraphics[width=12cm]{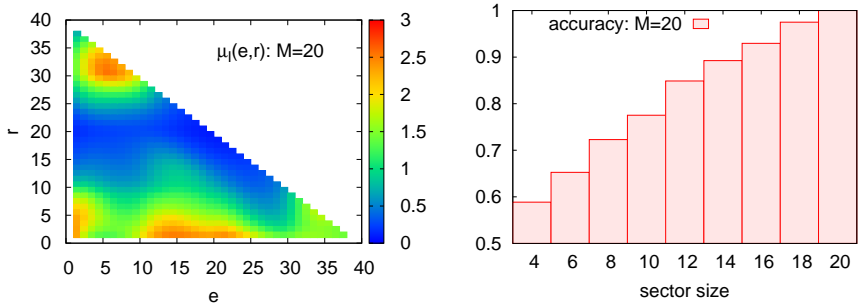} 
\caption{(left) The one-link probability distribution $\mu_l(e,r)$ obtained by the inverse algorithm given the average numbers $M^*(r),N_{p,s,x}^*(d)$ extracted from $10000$ randomly generated configurations of $M=20$ links in presence of a hard sector of size $L/2$ in the center of the chain. Here $(e,r)$ gives the first endpoint $e$ and length $r$ of the link. We note the presence of short link inside and outside the sector in addition to a subset of long links leaping over the sector. (right) Accuracy ($\#$ true positive + $\#$ true negative)/(total number of sites) of the inverse algorithm using the reconstructed one-link distribution $\mu_l(e,r)$ to infer regular sectors of different sizes in the center of the chain.}\label{f6}
\end{figure}

In practice, we solve the inverse problem by iteration \cite{Mac-book-2003,KR-neural-1998}: Starting from an initial set of parameters we compute the above averages within the Bethe approximation. We then apply incremental changes to the parameters according to deviations of the average values from their target values $M^*(r)$ and $N_{p,s,x}^*(d)$,
\begin{align}
\delta \lambda(r) &=\eta \big[ M^*(r)-\langle M(r) \rangle \big],\\
\delta \lambda_{p,s,x}(d) &=\eta \big[ N_{p,s,x}^*(d)-\langle N_{p,s,x}(d) \rangle \big],
\end{align}
with $0<\eta \ll 1$. Figure \ref{f6} shows the one-link probability distributions in the reconstructed model obtained, using this protocol, for a system of link variables in the presence of a hard sector. Here the sector consists of $L/2$ sites in the middle of the chain, and the data come from randomly generated link configurations respecting the constraints imposed by the sector (Fig \ref{f1} shows one sample configuration). As the figure shows, we can observe the signature of the sector already in the one-link distribution $\mu_l(e,r)$. Finally, we can repeat the above procedure to find better models, but this time we add an external field to disfavor the less probable connections suggested by $\mu_l(e,r)$ in the previous stage (see Appendix \ref{figs-app}). 

In principle, to infer the sector sites we need to study the likelihood of the model \cite{Mac-book-2003}, $\propto \mathrm{Pr}(\boldsymbol\sigma)\mathrm{Pr}(\mathcal{D}|\boldsymbol\lambda,\boldsymbol\sigma)$, where $\boldsymbol\sigma$ defines the position of the sector sites, and $\mathrm{Pr}(\boldsymbol\sigma)$ gives the prior probability of having $\boldsymbol\sigma$. $\mathrm{Pr}(\mathcal{D}|\boldsymbol\lambda,\boldsymbol\sigma)$ is the probability of observing the data $\mathcal{D}$ given the model parameters $\boldsymbol\lambda$ and the sector $\boldsymbol\sigma$. Here, for simplicity, we try a naive two-stage strategy using the reconstructed one-link probability distribution.       

More specifically, given the $\mu_l(e,r)$ we infer the sector contact sites by maximizing the probability 
\begin{align}
\mathcal{P}(\boldsymbol\sigma) \propto \prod_{i<j}[\alpha_{ij}]^{\sigma_i\sigma_j+(1-\sigma_i)(1-\sigma_j)}[1-\alpha_{ij}]^{1-\sigma_i\sigma_j-(1-\sigma_i)(1-\sigma_j)}\equiv e^{-\mathcal{E}(\boldsymbol\sigma)}.
\end{align}
Here $(i=e,j=e+r)$, and $\alpha_{ij}\equiv M\mu_l(i,j)(1-\mu_l(i,j))^{M/2-1}$ is the connection probability. Moreover, $\sigma_i=1$ if site $i$ belongs to the sector, otherwise $\sigma_i=0$. To find the $\boldsymbol\sigma$ maximizing the above probability, we make use of the Bethe approximation to compute the marginals $\nu_i(\sigma_i)$ of the joint probability distribution $\propto e^{-\beta \mathcal{E}(\boldsymbol\sigma)}$, where $\beta$ is an inverse temperature parameter. Then, we take the limit $\beta\to \infty$ of the finite-temperature BP equations to obtain the so-called minsum equations \cite{KFL-inform-2001}, by assuming an appropriate scaling for the BP marginals (see Appendix \ref{MS-app}),   
\begin{align}
h_{i \to j}=\sum_{k\neq i,j}\max(\ln(1-\alpha_{ik}),\ln \alpha_{ik}+h_{k\to i})
-\sum_{k\neq i,j}\max(\ln \alpha_{ik},\ln(1-\alpha_{ik})+h_{k\to i}).
\end{align}
We solve the equations for the cavity messages $h_{i \to j}$ by iteration, and compute the local messages $h_{i}$ considering all the incoming messages from the neighboring variables. These messages are used in a decimation or reinforcement algorithm \cite{BR-prl-2006} to find a good representative of the minimum energy configuration. In this way, we could infer the sector sites in a system of $M=20$ links with an accuracy, ($\#$ true positive + $\#$ true negative)/(total number of sites), that approaches $1$ for regular sectors of size $L/2$, see Fig. \ref{f6}. The accuracy is of course smaller for smaller (also for random) sectors where one needs more accurate inference algorithms considering the whole likelihood of the model to capture the subtle sector information in the data.

\section{Discussion}\label{S5}

Circuit topology is a molecular property that  describes arrangement of intra molecular contacts and critically determines dynamics and function of biomolecules such as proteins and nucleic acids \cite{MWT-structure-2014}. Here we explored the space of available circuit topologies through clarifying fundamental constraints on the configuration matrix $\mathsf{A}$. We then studied how one can construct a link configuration given an ordered configuration matrix. We showed that the problem of deciding on the validity of such a configuration matrix is computationally easy. 

We used the cavity method of statistical physics to explore the space of link configurations using an energy function of the two-link densities $n_{p,s,x}$. Exact computation of the entropy function $S(n_p,n_s)$ for small number of links shows that configurations with small $n_s$ are more frequent than those of small $n_p$. Approximate computations for larger number of links show similar behaviors for the entropy function. Moreover, the convergence pattern of the BP algorithm suggests that link variables are more correlated in the region of small $n_p$ than for small $n_s$ or $n_x$. This is close to the region that typical link configurations exhibit some degree of modularity, as observed for $\lambda_p=0,\lambda_{s,x}=1$ in Fig. \ref{f3}.  

The analysis of the one-link and two-link probability distributions showed that structure of link configurations changes considerably by varying the energy parameters $\lambda_{p,s,x}$ (conjugate to densities $n_{p,s,x}$); in particular, the typical configurations for $n_{s,x} > n_p$ exhibit a modular structure with components concentrated on the two sides of the chain. For the neutral choice of the energy parameters $\lambda_{p,s,x}=0$, two parallel or cross links are mostly found close to each other, whereas two series links are separated by a nonzero characteristic length scale $d_s^*$. The distance $d_s^*$ decreases by increasing $n_p$, where two parallel links are typically found at distance $d_p^*$. Similarly, increasing $n_x$ defines a nonzero length scale $d_x^*$ but this time $d_s^*$ increases.    

The constraints imposed by sectors of different sizes affect the statistical properties of the system in an unexpected way; in fact, for fixed energy parameters $\lambda_{p,s,x}$, the entropy could display a local maximum for an intermediate value of the sector size. 
Finally, we used the one- and two-link statistics in a learning algorithm to reconstruct the energy function that describes statistically the typical link configurations in presence of sectors. The information contained in the reconstructed model enables us to infer a meaningful number of the sector sites by a naive two-stage algorithm. In particular, we can successfully infer regular sectors of size $L/2$ in the center of chain.      

The simplicity of the model system used in this study serves in disentangling the role of topology from other structural features such as size, steric constraints and chemical structure of the polymer building blocks. In realistic settings, considering non-topological properties is often inevitable. We note that the energy constraint used in this study can be tailor-made for different molecular systems based on experimental data. Our aim in this article was to provide proof-of-concept study of using statistical mechanics to molecular topology. 

\section{Conclusion}
Structural topology of a linearly sorted multi-component system is often a critical determinant of its function and dynamics, thus controlling the topology of the system is of prime importance. In bimolecular systems, native topology reflects the evolutionary constraints imposed on the systems, while in engineering systems one needs to constraint the space of available topologies to ensure desired outcomes. For biomolecular system, arrangement of intramolecular contacts is the most relevant topological feature. Here, using a simple model of folded linear biomolecule, we explored the space of available contact arrangements and examined the impact of constraints on the topology of folded linear chains. Our approach enables identification of the underlying structural design principles and inference of associated evolutionary constraints and, as such, it potentially helps in understanding the functioning and the evolution of these systems. Further it may inspire engineers to build molecular systems with new functionalities for technological applications. 
 
Structural topology of a molecular system is often a critical determinant of its function and dynamics, thus controlling the topology of the system is of prime importance. For linear biopolymers, arrangement of intramolecular contacts is the most relevant topological feature. Here, using a simple model of folded linear biomolecules, we explored, for the first time, the space of available contact arrangements and examined the impact of constraints on the topology of folded linear chains. We found that the form of the imposed constraints critically determine not only the pairwise arrangement of contacts but also the global shape of the folded polymer. This capability allowed us to systematically search for the emergence of folded domains and sectors by varying the relevant order parameters. 
Our finding is particularly important for molecular engineering where one needs to constraint the space of available topologies to ensure desired functional outcomes. 

Our approach enables identification of the underlying structural design principles and inference of associated constraints and, as such, it potentially helps in understanding the functioning and the evolution of folded biomolecules, ranging from protein and RNA to chromosomes (e.g. topologically associated domains in chromosomes). 
In biomolecular systems, native topology reflects the constraints imposed on the systems during the synthesis as well as those imposed in the course of evolution. 
We demonstrated how one could infer the form of the imposed constraints from naturally occurring arrangement of contacts, which can in turn be readily extracted from coordinate files available in the databases. In the future, we foresee application of this approach to studies on molecular evolution as well on in vivo molecular folding processes where a number of physical constraints guide the conformational search of biopolymers to their native states. Further, the capability to infer constraints from natural systems may inspire engineers to build molecular systems with new functionalities for technological applications.

\acknowledgments
We would like to thank A. Mugler and R. Zecchina for their careful reading of our manuscript and their helpful comments.

\newpage
\appendix

\section{Details of the belief propagation equations}\label{BP-app}
We start from the partition function 
\begin{align}\label{Z}
Z(\lambda_p,\lambda_s,\lambda_x)=\frac{1}{M!}\sum_{\mathbf{e}}  \prod_{l<l'} \left( \delta_{e_l\neq e_{l'}}e^{\tilde{\lambda}_p\delta_{\mathsf{q}(e_l;e_{l'}),p}+\tilde{\lambda}_s\delta_{\mathsf{q}(e_l;e_{l'}),s}+\tilde{\lambda}_x\delta_{\mathsf{q}(e_l;e_{l'}),x} } \right),
\end{align}
which is a weighted sum over the link configurations $\mathbf{e}$ satisfying the perfect matching constraints. We divided the right hand side by $M!$ to cancel the overcounting resulted from different link permutations. 
For large $M$, the partition function can be rewritten as
\begin{multline}
Z(\lambda_p,\lambda_s,\lambda_x)=e^{(M\ln M) \phi(\lambda_p,\lambda_s,\lambda_x)}\\ \approx \int dn_p dn_s dn_x e^{(M\ln M)[S(n_p,n_s,n_x)+\lambda_p n_p+\lambda_s n_s+\lambda_x n_x]}.
\end{multline}
Here $e^{(M\ln M)S(n_p,n_s,n_x)}$ is the number of matchings of given densities $n_{p,s,x}=N_{p,s,x}/N$.
Note that the total number of perfect matchings is $e^{\ln (2M)!-\ln M!-M\ln 2}$ which for large $M$ scales as $e^{M\ln M-M(1-\ln 2)}$. 
Moreover, we have $\tilde{\lambda}_{p,s,x}=2\frac{\ln M}{M-1}\lambda_{p,s,x}$. 

We will use the Bethe approximation to compute $\phi(\lambda_p,\lambda_s,\lambda_x)$, and then by the Legendre transformation we will obtain the entropy function,
\begin{align}\label{sbp-app}
S(n_p^*,n_s^*,n_x^*) =\phi(\lambda_p,\lambda_s,\lambda_x)-\lambda_p n_p^*-\lambda_s n_s^*-\lambda_x n_x^*.
\end{align}
The values $n_{p,s,x}^*$ are determined by the saddle-point equations, 
\begin{align}
n_p^* =\langle n_p \rangle=\frac{\partial \phi}{\partial \lambda_p}=\frac{2}{M(M-1)} \sum_{l<l'}\langle \delta_{\mathsf{q}(e_l,e_{l'}),p} \rangle,
\end{align}
and similarly for $n_{s,x}^*$.

The central quantities in the Bethe approximation are the cavity marginals $\mu_{l\to l'}(e_l)$, giving the probability of having endpoints $e_l$ for link $l$ in the absence of interactions and constraints involving $e_{l'}$ \cite{MM-book-2009,KFL-inform-2001}. The cavity marginal $\mu_{l\to l'}(e_l)$ is obtained by considering the cavity messages from the other variables $\mu_{l''\to l}(e_{l''})$, and the local constraints depending on the $(e_{l},e_{l''})$,       
\begin{align}\label{bp-app}
\mu_{l\to l'}(e_l) \propto \prod_{l'' \ne l,l'} \left( \sum_{e_{l''} \ne e_l} e^{\tilde{\lambda}_p\delta_{\mathsf{q}(e_l;e_{l''}),p}+\tilde{\lambda}_s\delta_{\mathsf{q}(e_l;e_{l''}),s}+\tilde{\lambda}_x\delta_{\mathsf{q}(e_l;e_{l''}),x}} \mu_{l'' \to l}(e_{l''})\right).
\end{align}
We solve the equations by iteration, starting from random initial marginals and updating the $\mu_{l\to l'}(e_l)$ in a random sequential way according to the above equations. 

Having the cavity marginals, the two-link marginals read
\begin{align}\label{bp2-app}
\mu_{l,l'}(e_l,e_{l'}) \propto  \delta_{e_{l'} \ne e_l} e^{\tilde{\lambda}_p\delta_{\mathsf{q}(e_l;e_{l'}),p}+\tilde{\lambda}_s\delta_{\mathsf{q}(e_l;e_{l'}),s}+\tilde{\lambda}_x\delta_{\mathsf{q}(e_l;e_{l'}),x}} \mu_{l \to l'}(e_{l})\mu_{l' \to l}(e_{l'}).
\end{align}
The free energy in the Bethe approximation is given by $(M\ln M)\phi= \sum_{l} \Delta \phi_l-\sum_{l<l'} \Delta \phi_{ll'}-\ln M!$, where $\Delta \phi_l$ and $\Delta \phi_{ll'}$ are the local free energy shifts \cite{MM-book-2009}. These are the changes in the free energy after adding the constraints and the energy terms depending on $e_l$, and those that involve $(e_{l},e_{l'})$, namely, 
\begin{align}\label{fbp1-app}
e^{\Delta \phi_l} &= \sum_{e_l}\prod_{l' \ne l} \left( \sum_{e_{l'} \ne e_l} e^{\tilde{\lambda}_p\delta_{\mathsf{q}(e_l;e_{l'}),p}+\tilde{\lambda}_s\delta_{\mathsf{q}(e_l;e_{l'}),s}+\tilde{\lambda}_x\delta_{\mathsf{q}(e_l;e_{l'}),x}} \mu_{l' \to l}(e_{l'})\right),\\ \label{fbp2-app}
e^{\Delta \phi_{ll'}} &= \sum_{e_l \ne e_{l'}} e^{\tilde{\lambda}_p\delta_{\mathsf{q}(e_l;e_{l'}),p}+\tilde{\lambda}_s\delta_{\mathsf{q}(e_l;e_{l'}),s}+\tilde{\lambda}_x\delta_{\mathsf{q}(e_l;e_{l'}),x}} \mu_{l \to l'}(e_{l})
\mu_{l' \to l}(e_{l'}).
\end{align}

Here the links are equivalent, so we rewrite the BP equations as
\begin{align}\label{ubp-app}
\mu(e_l) \propto \left( \sum_{e_{l'} \ne e_l} e^{\tilde{\lambda}_p\delta_{\mathsf{q}(e_l;e_{l'}),p}+\tilde{\lambda}_s\delta_{\mathsf{q}(e_l;e_{l'}),s}+\tilde{\lambda}_x\delta_{\mathsf{q}(e_l;e_{l'}),x}} \mu(e_{l'})\right)^{M-2}.
\end{align}
Let us represent $e_l$ by its first endpoint and its length $(e,r)$, then the above equation reads
\begin{align}
\mu(e,r) \propto [ e^{\tilde{\lambda}_p} w_p(e,r)+e^{\tilde{\lambda}_s} w_s(e,r)+e^{\tilde{\lambda}_x} w_x(e,r)]^{M-2}.
\end{align}
where
\begin{align}
w_p(e,r) &=\sum_{e'=1}^{e-1}\sum_{r'=e+r+1-e'}^{2M-e'} \mu(e',r')+ \sum_{e'=e+1}^{e+r-2}\sum_{r'=1}^{e+r-1-e'}  \mu(e',r'),\\
w_s(e,r) &= \sum_{e'=1}^{e-2}\sum_{r'=1}^{e-e'-1} \mu(e',r')+ \sum_{e'=e+r+1}^{2M-1}\sum_{r'=1}^{2M-e'}  \mu(e',r'), \\
w_x(e,r) &=\sum_{e'=1}^{e-1}\sum_{r'=e-e'+1}^{e+r-1-e'} \mu(e',r')+ \sum_{e'=e+1}^{e+r-1}\sum_{r'=e+r+1-e'}^{2M-e'}  \mu(e',r').
\end{align}
Similarly, we obtain 
\begin{align}
e^{\Delta \phi_l} &= \sum_{e=1}^{2M-1}\sum_{r=1}^{2M-e} [ e^{\tilde{\lambda}_p} w_p(e,r)+e^{\tilde{\lambda}_s} w_s(e,r)+e^{\tilde{\lambda}_x} w_x(e,r)]^{M-1},\\
e^{\Delta \phi_{ll'}} &=\sum_{e=1}^{2M-1}\sum_{r=1}^{2M-e} \mu(e,r)[ e^{\tilde{\lambda}_p} w_p(e,r)+e^{\tilde{\lambda}_s} w_s(e,r)+e^{\tilde{\lambda}_x} w_x(e,r)].
\end{align}
Moreover, we have
\begin{align}
\langle n_p \rangle =e^{-\Delta \phi_{ll'}}e^{\tilde{\lambda}_p}\sum_{e=1}^{2M-1}\sum_{r=1}^{2M-e} \mu(e,r)w_p(e,r).
\end{align}

\subsection{An alternative representation}
We may as well use the matching variables $c_{ij}\in \{0,1\}$, showing the connectivity of nodes $i$ and $j$, to rewrite the partition function \ref{Z} as
\begin{align}\label{Zc}
Z(\lambda_p,\lambda_s,\lambda_x)=\sum_{\mathbf{c}}  \prod_{i} \delta_{\sum_{j\ne i}c_{ij}=1} \prod_{(ij)<(kl)}e^{c_{ij}c_{kl}[\tilde{\lambda}_p\delta_{\mathsf{q}(ij;kl),p}+\tilde{\lambda}_s\delta_{\mathsf{q}(ij;kl),s}+\tilde{\lambda}_x\delta_{\mathsf{q}(ij;kl),x}] },
\end{align}
This representation of the problem is more efficient than the one we used above but the BP equations are more involved; we have to distinguish between two kinds of BP message $\mu_{(ij) \to i}(c_{ij})$ and $\mu_{(ij) \to (kl)}(c_{ij})$. The former is the probability of $c_{ij}$ in absence of the matching constraint $I_i(c_{\partial i})\equiv \delta_{\sum_{j\ne i}c_{ij}=1}$, and the latter is computed in absence of two-link interaction $w_{(ij),(kl)}(c_{ij},c_{kl})\equiv \exp(c_{ij}c_{kl}[\tilde{\lambda}_p\delta_{\mathsf{q}(ij;kl),p}+\tilde{\lambda}_s\delta_{\mathsf{q}(ij;kl),s}+\tilde{\lambda}_x\delta_{\mathsf{q}(ij;kl),x}])$. Here $c_{\partial i}\equiv \{c_{ij}: j \neq i\}$. The BP equations governing these cavity marginals are
\begin{multline}
\mu_{(ij) \to j}(c_{ij})\propto \left( \sum_{c_{\partial i\setminus j}}I_i(c_{\partial i})\prod_{k\in \partial i\setminus j}\mu_{(ik)\to i}(c_{ik}) \right) \\ \times
\prod_{(kl):k,l\neq i,j} \left(\sum_{c_{kl}}w_{(ij),(kl)}(c_{ij},c_{kl})\mu_{(kl)\to (ij)}(c_{kl})\right),
\end{multline}
and
\begin{multline}
\mu_{(ij) \to (kl)}(c_{ij})\propto \left( \sum_{c_{\partial i\setminus j}}I_i(c_{\partial i})\prod_{k\in \partial i\setminus j}\mu_{(ik)\to i}(c_{ik}) \right)
\left( \sum_{c_{\partial j\setminus i}}I_j(c_{\partial j})\prod_{k\in \partial j\setminus i}\mu_{(jk)\to j}(c_{jk}) \right) \\ \times
\prod_{(k'l')\neq (kl): k',l'\neq i,j} \left(\sum_{c_{k'l'}} w_{(ij),(k'l')}(c_{ij},c_{k'l'})\mu_{(k'l')\to (ij)}(c_{k'l'})\right).
\end{multline}

Similarly, we can compute the one-link marginals $\mu_{(ij)}(c_{ij})$ and the two-link marginals $\mu_{(ij),(kl)}(c_{ij},c_{kl})$.

\section{Details of the minsum equations}\label{MS-app}
Consider a system of interacting site variables $\sigma_i\in \{0,1\}$ for $i=1,\dots,L$, with energy function 
$\mathcal{E}(\boldsymbol\sigma)=\sum_{i<j}\mathcal{E}_{ij}(\sigma_i,\sigma_j)$, 
where
\begin{multline}
\mathcal{E}_{ij}(\sigma_i,\sigma_j)=- [\sigma_i\sigma_j+(1-\sigma_i)(1-\sigma_j)]\ln \alpha_{ij}\\
-[1-\sigma_i\sigma_j-(1-\sigma_i)(1-\sigma_j)]\ln(1-\alpha_{ij}).
\end{multline}
The $\alpha_{ij}$ are here parameters, giving the probability of having a link connecting sites $i,j$.

We start from the finite-temperature BP equations for the cavity marginals of the probability distribution of variable configurations $\mathcal{P}(\boldsymbol\sigma) \propto e^{-\beta \mathcal{E}(\boldsymbol\sigma)}$,
\begin{align}
\nu_{i\to j}(\sigma_i) \propto \prod_{k \ne i,j} \left( \sum_{\sigma_k} e^{-\beta \mathcal{E}_{ik}(\sigma_i,\sigma_k)} \nu_{k \to i}(\sigma_k)\right).
\end{align} 
This is the probability of state $\sigma_i$ for site $i$ in the absence of interaction with site $j$.
It is more appropriate to work with the cavity fields $h_{i\to j}\equiv \frac{1}{\beta}\ln\left( \frac{\nu_{i\to j}(1)}{\nu_{i\to j}(0)}\right)$ where the BP equations read
\begin{multline}
\beta h_{i\to j}= \sum_{k \ne i,j}\ln\left( e^{-\beta\mathcal{E}_{ik}(1,0)}+ e^{-\beta\mathcal{E}_{ik}(1,1)+\beta h_{k \to i}(\sigma_k)}\right)\\
-\sum_{k \ne i,j}\ln\left( e^{-\beta\mathcal{E}_{ik}(0,0)}+ e^{-\beta\mathcal{E}_{ik}(0,1)+\beta h_{k \to i}(\sigma_k)}\right).
\end{multline}

Now take the limit $\beta \to \infty$ of the above equations.
The resulting equations for the cavity messages $h_{i \to j}$ are called minsum equations \cite{KFL-inform-2001} and read
\begin{align}
h_{i \to j}=\sum_{k\neq i,j}\max(\ln(1-\alpha_{ik}),\ln \alpha_{ik}+h_{k\to i})
-\sum_{k\neq i,j}\max(\ln \alpha_{ik},\ln(1-\alpha_{ik})+h_{k\to i}).
\end{align}
We solve the minsum equations for the cavity messages by iteration, starting from random initial messages. In the end, the local messages $h_{i}$ are obtained like the cavity ones but considering all the incoming messages from the neighboring variables.  

To find a configuration minimizing the energy, we use the reinforcement algorithm \cite{BR-prl-2006}: In each step of updating the cavity messages, we add external fields that polarize the messages in the direction suggested by the local messages. More precisely, the reinforced minsum equations read 
\begin{multline}
h_{i \to j}^{t+1}=r(t)h_{i}^t+\sum_{k\neq i,j}\max(\ln(1-\alpha_{ik}),\ln \alpha_{ik}+h_{k\to i}^t)\\
-\sum_{k\neq i,j}\max(\ln \alpha_{ik},\ln(1-\alpha_{ik})+h_{k\to i}^t).
\end{multline}
Similarly, we update the local messages
\begin{multline}
h_{i}^{t+1}=r(t)h_{i}^t+\sum_{k\neq i}\max(\ln(1-\alpha_{ik}),\ln \alpha_{ik}+h_{k\to i}^t)\\
-\sum_{k\neq i}\max(\ln \alpha_{ik},\ln(1-\alpha_{ik})+h_{k\to i}^t).
\end{multline}
The reinforcement parameter $r(t)$ is zero at the beginning of the algorithm ($t=0$) and grows slowly by $t$, for example as $r(t+1)=r(t)+\delta$.    

\section{More details and figures}\label{figs-app}
Here we present more details of the numerical simulations and figures obtained in this study.
  
\begin{figure}
\includegraphics[width=15cm]{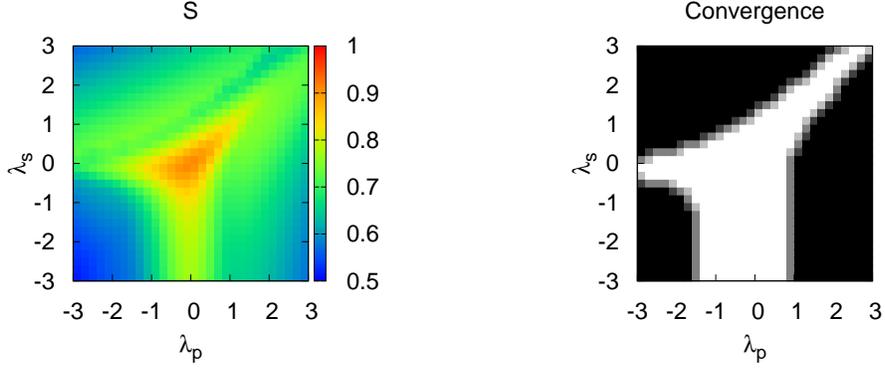} 
\caption{The entropy $S=\frac{1}{M\ln M}\ln \mathcal{N}$ ($\mathcal{N}$ is the number of configurations) obtained by the Bethe approximation, and the region that the BP algorithm converges (white region). The data are for $M=40$ links and $\lambda_x=0$. The convergence limit is  $\epsilon=10^{-8}$.}\label{f1-app}
\end{figure}

Figure \ref{f1-app} displays the approximate entropy (logarithm of the number of link configurations $\mathcal{N}$) that is obtained within the Bethe approximation. Here we take $\lambda_x=0$ and report the entropy in the space of parameters $(\lambda_p,\lambda_s)$ even if the algorithm does not converges. The solution to the BP equations \ref{ubp-app} is found by iteration and the algorithm converges when the difference in the BP messages $\mu(e_l)$ in two successive steps of the iteration is less than a convergence limit $\epsilon=10^{-8}$. In the same figure, we observe the region in the parameter space that the BP algorithm converges. 
Given the BP messages, the entropy is computed by Eqs. \ref{sbp-app} and \ref{fbp1-app},\ref{fbp2-app}.    

\begin{figure}
\includegraphics[width=15cm]{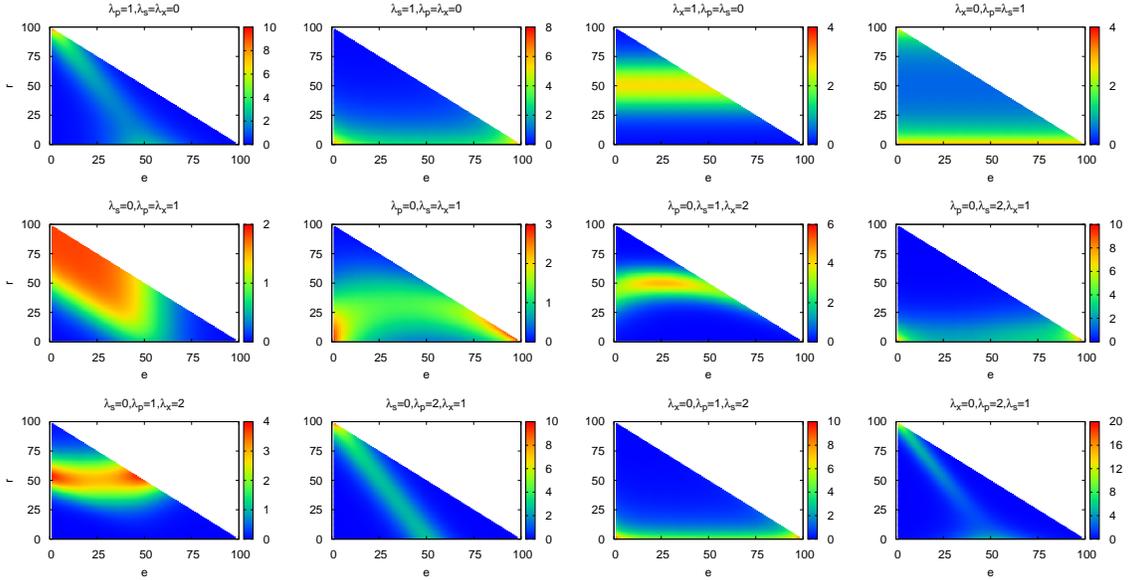} 
\caption{One-link distribution (more precisely $M(2M-1)\mu(e,r)$) obtained by the Bethe approximation for different energy parameters $\lambda_{p,s,x}$ with $M=50$ links. Here $\mu(e,r)$ is the probability of having a link with the first endpoint $e$ and length $r$.}\label{f2-app}
\end{figure}

\begin{figure}
\includegraphics[width=15cm]{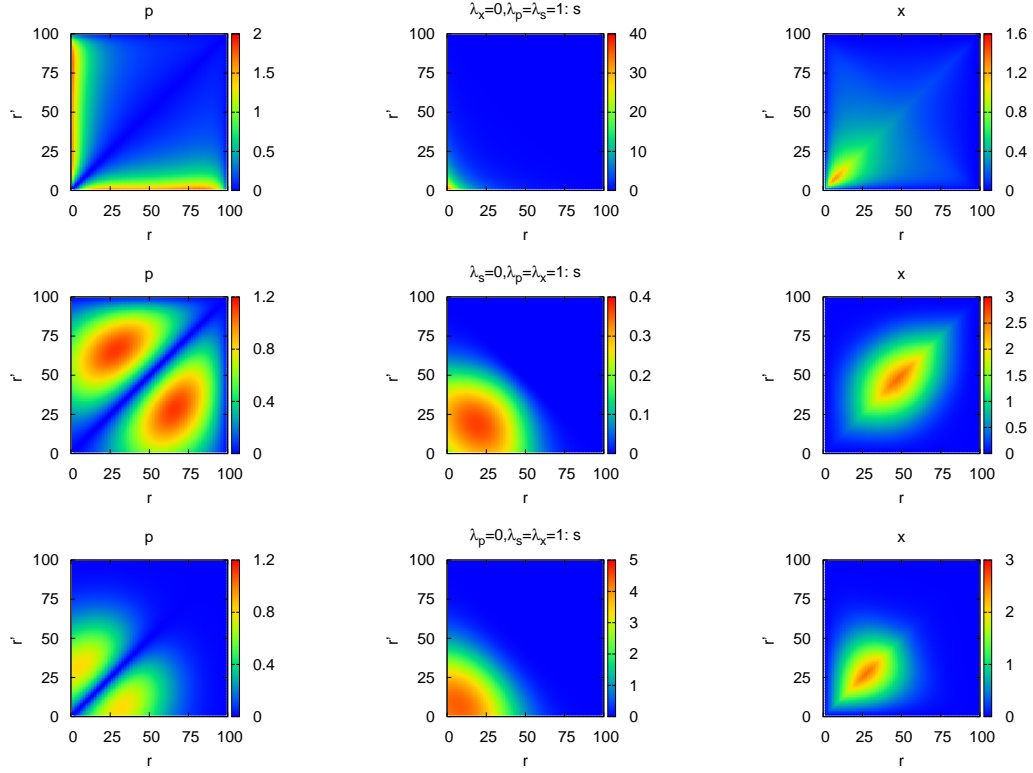} 
\caption{Two-link length distribution $\mu_{ll'}(r,r')$ (multiplied by a constant to make it of order one) obtained by the Bethe approximation for $M=50$ links. Here $\mu_{ll',q}(r,r')$ is the probability of finding two-links of type $q=p,s,x$ with lengths $r$ and $r'$. The energy parameters $\lambda_{p,s,x}$ and the average two-link densities are fixed in each row. The columns are for different types of two-links: $p$ (left), $s$ (center), and $x$ (right).}\label{f3-app}
\end{figure}

\begin{figure}
\includegraphics[width=15cm]{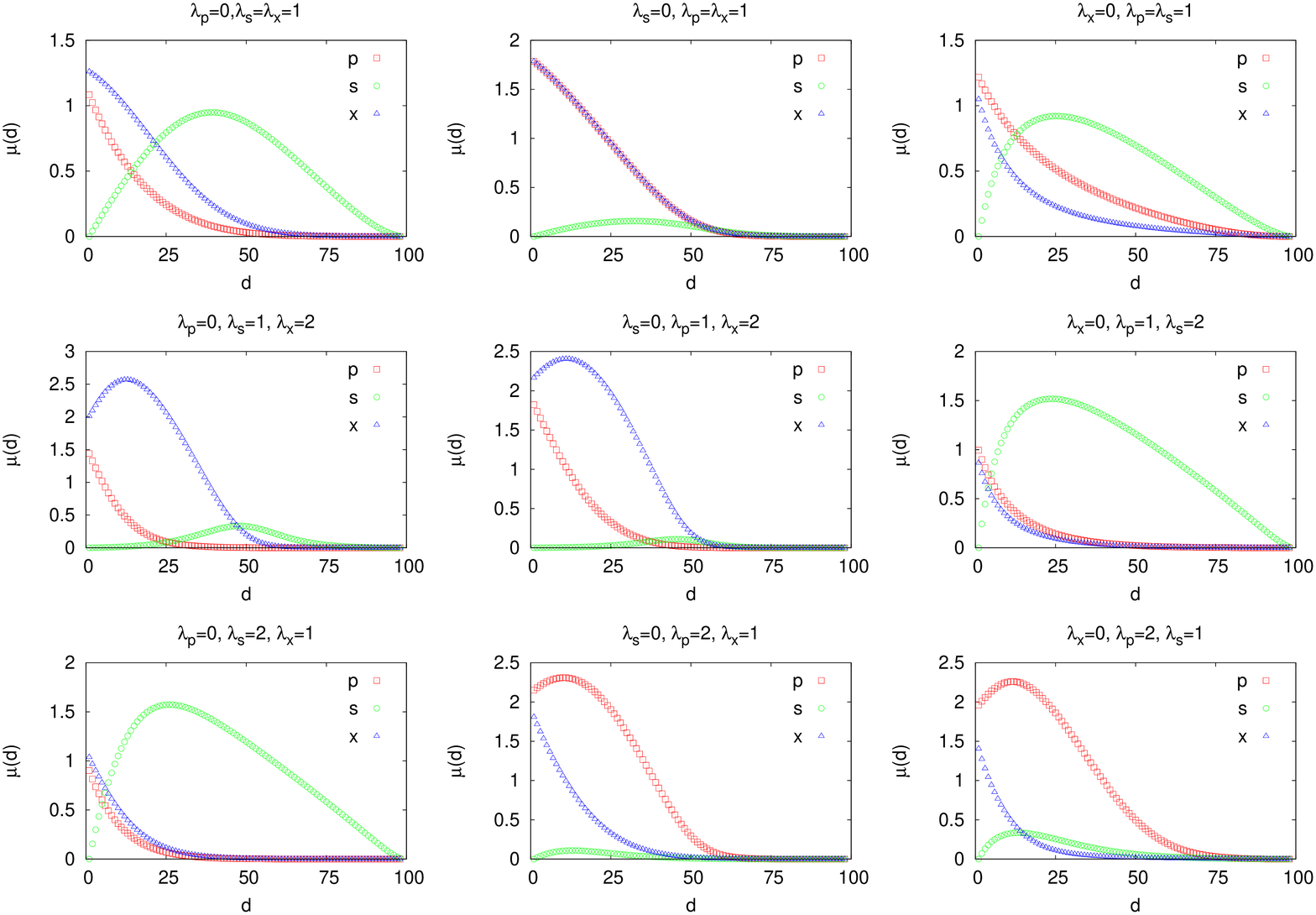} 
\caption{Two-link distance distribution $\mu_{ll'}(d)$ (multiplied by a constant to make it of order one) for different energy parameters $\lambda_{p,s,x}$ and different types of two-links $(p,s,x)$, obtained by the Bethe approximation for $M=50$ links. Here $\mu_{ll',q}(d)$ is the probability of finding two links of type $q=p,s,x$ at distance $d$ (separation of the first endpoints) from each other.}\label{f4-app}
\end{figure}

\begin{figure}
\includegraphics[width=16cm]{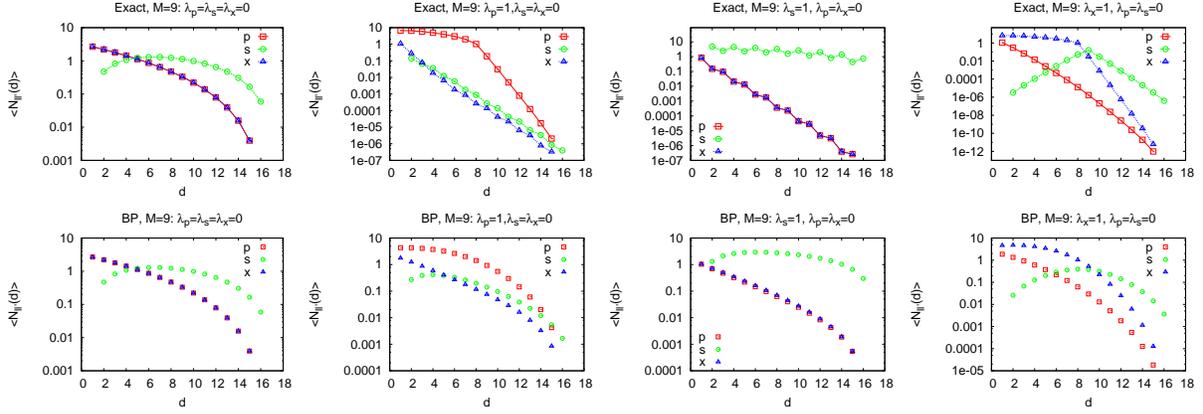} 
\caption{Comparing the average two-link numbers $\langle N_{ll'}(d)\rangle$ obtained exactly (top) with those of the Bethe approximation (bottom) for $M=9$ links. Distance $d$ of two links is the separation of their first endpoints. Each panel shows $\langle N_{ll'}(d)\rangle$ for fixed energy parameters $\lambda_{p,s,x}$ but different types $p,s,x$.}\label{f5-app}
\end{figure}

Figures \ref{f2-app}, \ref{f3-app}, and \ref{f4-app} show the one-link and two-link distributions for more parameter samples obtained by the BP algorithm as described above. In Fig. \ref{f5-app}, we compare the two-link distance distribution obtained by the approximate algorithm with the exact one for a small number of links.

\begin{figure}
\includegraphics[width=12cm]{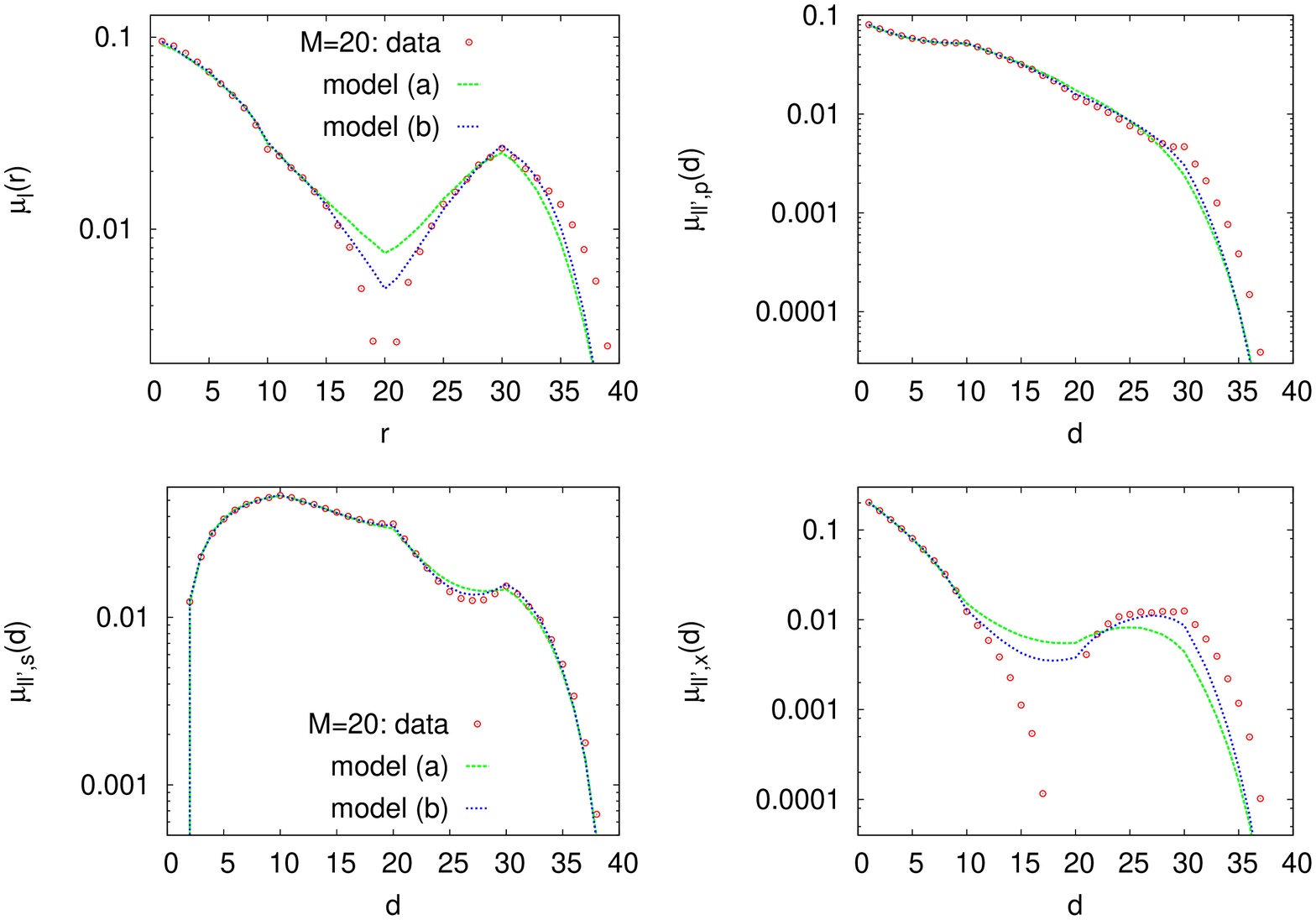} 
\caption{The one- and two-link probability distributions $\mu_l(r), \mu_{ll',p,s,x}(d)$ obtained by the inverse algorithm given the average numbers $M^*(r),N_{p,s,x}^*(d)$ extracted from $10000$ randomly generated configurations of $M=20$ links in presence of a regular sector of size $L/2$ in the center of the chain. Hear $r$ refers to the length of link, and $d$ gives the distance between the first endpoints of two links. $M^*(r)$ and $N_{p,s,x}^*(d)$ are the average number of links of length $r$ and two-links of distance $d$, respectively. Model (a) is obtained by running the inverse algorithm with no prior information of the sector. To obtain model (b), we run the inverse algorithm with an additional external field disfavoring some connections according to the one-link probability distribution $\mu_l(e,r)$ provided by model (a). $\mu_l(e,r)$ is the probability of having a link with first endpoint $e$ and length $r$.}\label{f6-app}
\end{figure}

In Fig. \ref{f6-app}, we compare the reconstructed one-link and two-link distributions with the observed data from link configurations    
with a regular sector of size $L/2$. The inferred statistics can be improved by iteration using the information obtained in the previous stages of the algorithm. In the figure, we also compare the model data obtained without any prior information (a), and with the information provided in the first stage of the algorithm (b).

\begin{thebibliography}{prsty}

\bibitem{MWT-structure-2014} A. Mashaghi, R.J. van Wijk and S.J. Tans, Structure {\bf 22}(9), 1227?1237 (2014)

\bibitem{MTM-pccp-2014} A. Mugler, S.J. Tans and A. Mashaghi,  Phys Chem Chem Phys. {\bf 16}(41):22537-44 (2014)

\bibitem{H-embo-1986} G. von Heijne, EMBO 5(11): 3021?3027 (1986)

\bibitem{H-nature-2006} G. von Heijne, Nature Reviews Molecular Cell Biology 7, 909-918 (2006)

\bibitem{MGV-epl-1986}M. Mezard, G. Parisi, and M. A. Virasoro,  Europhys. Lett. \textbf{1}, 77 (1986).

\bibitem{MP-epjb-2001} M. M\'ezard and G. Parisi,  Eur. Phys. J. B \textbf{20}, 217 (2001).

\bibitem{MM-book-2009} M. M\'ezard and A. Montanari, {\it Information, Physics, and Computation} (Oxford University Press, Oxford, 2009).

\bibitem{MZ-pre-2002}M. M\'ezard and R. Zecchina, Phys. Rev. E \textbf{66}, 056126 (2002).

\bibitem{MPZ-science-2002}M. M\'ezard, G. Parisi and R. Zecchina, Science \textbf{297}, 812 (2002).



\bibitem{ZM-jstat-2006}L. Zdeborova and M. Mezard,  J. Stat. Mech. (2006) P05003.

\bibitem{MS-jstat-2006}E. Marinari and G. Semerjian, J. Stat. Mech. (2006) P06019.

\bibitem{RZ-pre-2012}A. Ramezanpour and R. Zecchina, Phys. Rev. E \textbf{85}, 021106 (2012)

\bibitem{BBCZ-jstat-2008}M. Bayati, C. Borgs, J. Chayes and R. Zecchina, J. Stat. Mech. (2008) L06001.


\bibitem{HRLR-cell-2009} N. Halabi, O. Rivoire, S. Leibler, and R. Ranganathan,  Cell \textbf{138}: 774–786, (2009). 


\bibitem{KFL-inform-2001}F. R. Kschischang, B. J. Frey, and H. -A. Loeliger, IEEE Trans. Infor. Theory \textbf{47}, 498 (2001).

\bibitem{BR-prl-2006}A. Braunstein and R. Zecchina, Phys. Rev. Lett. \textbf{96}, 030201 (2006).


\bibitem{WWSHH-pnas-2009} M. Weigt, R. A. White, H. Szurmant, J. A. Hoch, and T. Hwa, Proc. Natl. Acad. Sci. USA \textbf{106} (1) 67, (2009).

\bibitem{PZS-plos-2011} D. S. Marks, L. J. Colwell, R. Sheridan, T. A. Hopf, A. Pagnani, R. Zecchina, and C. Sander, PLoS ONE \textbf{6}(12): e28766, (2011).
 
\bibitem{PZW-pnas-2011} F. Morcos, A. Pagnani, B. Lunt, A. Bertolino, D. S. Marks, C. Sander, R. Zecchina, J. N. Onuchic, T. Hwa, and  M. Weigt.
Proc. Natl. Acad. Sci. USA \textbf{108}: E1293–E1301, 2011.

\bibitem{BR-arxiv-2007}  W. Bialek, and R. Ranganathan, arXiv:0712.4397, (2007).


\bibitem{Mac-book-2003} D. J. C. MacKay, \textit{Information Theory, Inference, and Learning
Algorithms} (Cambridge University Press, Cambridge, U.K., 2003).
 
\bibitem{KR-neural-1998} H. J. Kappen, and F. B. Rodriguez,   Neural Computation \textbf{10}: 1137, (1998). 



\end{thebibliography}
\end{document}